\documentclass[12pt, a4paper]{article} 
\usepackage{verbatim}
\usepackage{amsmath}
\usepackage{amssymb}

\usepackage[german,english]{babel} 
\usepackage{booktabs} 
\usepackage{comment} 
\usepackage[utf8]{inputenc} 
\usepackage[T1]{fontenc} 

\usepackage{amsmath,amsfonts,amsthm} 
\usepackage{tikz} 
\usepackage{tikz-cd}
\usetikzlibrary{positioning,arrows} 
\usetikzlibrary{decorations.pathreplacing} 
\usepackage{tikzsymbols} 
\usepackage{blindtext} 

\usepackage{float}
\usepackage{natbib}
\usepackage{lscape}
\usepackage{booktabs}
\usepackage{graphicx}
\usepackage{caption}
\usepackage{adjustbox}

\usepackage[nottoc, notlot, notlof]{tocbibind}

\usepackage{geometry} 

\usepackage{setspace} 
\usepackage[T1]{fontenc} 
\usepackage[utf8]{inputenc} 

\usepackage{fancyhdr} 
\usepackage{natbib} 
\usepackage{har2nat} 

\usepackage{xcolor}
\usepackage[normalem]{ulem}
\useunder{\uline}{\ul}{}
\usepackage{longtable}
\usepackage{longtable}
\usepackage{array}
\usepackage{booktabs}
\usepackage{lscape}
\usepackage{geometry}
\usepackage{graphicx}
\usepackage{subcaption} 
\usepackage{amssymb} 

\title{TITLE} 
\author{
FIRSTNAME LASTNAME 
  }

\date{\small \today} 

\begin{document}





\begin{titlepage} 


\vspace*{1in} 
	
\center 

	
\textbf{\large\bfseries Which Identities Are Mobilized: Towards an automated detection of social group appeals in political texts}\\[0.4cm] 

{\large\today}\\[0.4cm] 
Felicia Riethmüller\\
Julian Dehne\\
Denise Al-Gaddooa
	
\vfill 


\vfill 

Prepared for the COMPTEXT Conference, 2-4 May 2024, Amsterdam \\ 
\textcolor{red}{Early draft! Please do not cite or circulate without the author's permission!} \\ 

\vfill 


\vfill 
	

\end{titlepage} 


\begin{abstract}

This paper proposes a computational text classification strategy to identify references to social groups in European party manifestos and beyond. Our methodology uses machine learning techniques, including BERT and large language models, to capture group-based appeals in texts. We propose to combine automated identification of social groups using the Mistral-7B-v0.1 Large Language Model with Embedding Space-based filtering to extend a sample of core social groups to all social groups mentioned in party manifestos. By applying this approach to RRP's and mainstream parties' group images in manifestos, we explore whether electoral dynamics explain similarities in group appeals and potential convergence or divergence in party strategies. Contrary to expectations, increasing RRP support or mainstream parties' vote loss does not necessarily lead to convergence in group appeals. Nonetheless, our methodology enables mapping similarities in group appeals across time and space in 15 European countries from 1980 to 2021 and can be transferred to other use cases as well.

\end{abstract} 
\setcounter{page}{1}
\section{Introduction}

The ascent of radical right Parties (RRPs) in European party systems has entailed an academic debate about mainstream and niche/challenger parties' strategies \cite{meguid2005competition, meguid2008party, meyer2013niche, abou2016niche, abou2020causal, vries2020political, bergman2020issue}. As a niche strategy generally means stressing a smaller and different set of issues than mainstream competitors, parties can for instance switch in and out of a niche issue profile \cite{wagner2012defining}. In the niche party literature, but also in the literature that defines nicheness as a strategy of challenger parties \cite{vries2020political}, parties' electoral fortunes are generally discussed based on issue salience or position-taking \cite{abou2016niche}. Thus, many studies have looked into whether there is positional or thematic accommodation of the Radical Right by mainstream parties \cite{bale2003cinderella, van2010contagious, abou2016niche, abou2020parties, gessler2022refugee}. Yet, although we see realignment processes on the voter side \cite{bornschier2021us}, findings on whether Radical Right and mainstream parties actually speak to the same social groups and voter segments are scarce.

That is why a recent argument holds that the current tripolar dynamics of party competition need a group identity perspective in order to fully understand the declines of mainstream- and the successes of challenger competitors. Next to their issue profile, parties also have a "group image" (Thau 2023, p.3)\nocite{thau2023group}, i.e. "public perceptions about which groups a party represents", which parties intend to shape in their favour. Of course, they can do so via policies, however, another powerful strategy lies in using direct group-based appeals \cite{dolinsky2023parties, huber_dolinsky_2023, huber2022beyond, thau2019political, stuckelberger2022group}. As group identity is a crucial factor moderating party attachment \cite{bornschier2021us, achen2016democracy, dickson2006social}, parties have an additional vehicle of competition in which groups they address and relate themselves to. With increasing competition between mainstream and (Radical Right) challenger parties, one should expect interactions between their group images. Yet, these have not been explored, particularly because a comprehensive assessment of all groups addressed across parties and countries has been lacking.

Despite these important results, researchers still face challenges identifying group-based rhetoric, especially in large-scale cross-national endeavours. It has proven especially difficult to identify the baseline/basic population of social groups in political texts instead of a predefined set of groups. Yet, identifying all groups mentioned in political texts is important for two reasons: 1) \textit{the constructed nature of social groups}: in principle all groupings of people can become a relevant social group identity in political discourse \cite{saward2006representative, zollinger2024cleavage}, i.e. parties have an active role in politicizing certain groups, and 2) \textit{semantic nuances}: it is likely that different parties address the same social groups differently, priming certain parts of their identities in order to reconcile their ideological profile with electoral concerns, as in the case of Germany's radical right AfD, which appeals to 'German women' or 'traditional women' instead of just 'women'. 

This is why we propose a computational text classification strategy to identify references to social groups in political texts. Group-based appeals are "intentional act[s] that associate a political actor with or dissociates them from a social group (Dolinsky and Huber 2023, p.11). A frequent form these group appeals take is written statements, which is why we use party manifestos from the Manifesto Corpus (Lehmann et al. 2023) to capture parties' group images. We apply three different identification methods, compare their performance and provide a strategy to combine their respective benefits for the task of comparatively identifying who political parties talk about in their manifestos.  The first method is a supervised identification via a dictionary of a set of social groups defined by the researcher. We use those groups that are implied in the sociodemographic variables of the CSES waves 1-5 \nocite{CSESIMD, CSESModule5}, to allow for matches between the supply- and demand side. This creates a set of "seed groups", which are then in a second step used to train a BERT \cite{devlin2019bert} multilingual language model. This model learns the semantic meaning of the social groups and can a) transfer these meanings to other languages and b) find alternative wordings that are semantically close but not included in the dictionary. This addresses the \textit{semantic nuances} problem of social group identification. The \textit{constructed nature of social groups} is more difficult to address for a BERT model trained on a set of specific groups, because inferring all \textit{possible} groups from these is quite a large step. That is why in a third step, we use the mistralai/Mistral-7B-v0.1 large language model, which allows us to input a manifesto sentence and a minimal definition of a social group \cite{licht_sczepanski_2023} and have the LLM output all explicitly mentioned social groups, without predefining these. To deal with the noise in the inference output, we propose the addition of Embedding Space-based filtering (ESF), which uses the embeddings of the seed groups and several classifiers to determine the optimal boundary of what constitutes a social group. 
Finally, we compare the group-based appeals of each of these three methods. Our combination of the \textit{Mistral LLM-identification and Embedding Space-based Filtering (LLM-ESF)} allows us to considerably enhance the "seed" groups defined in the original dictionary with minimal manual coding effort and thus to approach a tentative full sample of social groups mentioned in European party manifestos from 1980 to 2021. We thereby also provide an assessment of how well they can be identified by different methods of (un-)supervised text classification.

Existing approaches have mostly used either manual coding \nocite{huber2022beyond, dolinsky2023parties, thau2021social} (see e.g. Huber 2022, Dolinsky 2022, Thau 2021) or dictionaries \cite{riethmuller2024supply} to identify references to social groups. The most far-reaching approach to capturing group appeals in their linguistic diversity without predefining relevant groups comes from Licht and Sczepanski \citeyear{licht_sczepanski_2023}. The authors use a Transformer-based supervised token classifier to identify spans containing group mentions in texts. This is a very promising and useful approach. While Licht and Sczepanski's approach has the advantage of granting more oversight regarding which groupings are considered a social group to the researcher, our alternative strategy has the advantage of requiring very low manual coding efforts.

We put our computational group identification to the test and apply it to the comparison of RRPs and mainstream parties' group images. Whereas we know that accomodation of radical right niche issues by mainstream parties and, vice versa, issue diversification by radical right parties are strategies resulting from the radical rights' electoral success, we do not know yet whether mainstream and challenger parties actually target the same electorates and what causes convergence and divergence in group appeal strategies, i.e., in party's "group images" \cite{thau2023group}. We test whether electoral dynamics can explain similarity in group images, similar to contagion effects in spatial and issue competition. While we did not find that increasing Radical Right support or mainstream parties' vote loss makes Centre Left/Right and Radical Right parties move closer together in their group images, our approach allows us to map similarity in group appeals across time and space, namely in 15 countries from 1980 to 2021.


\section{Radical right parties' contested group image}

Who does the Radical Right stand for? At the center of this inquiry has been thus far the Radical Right's relationship to the working class, which has thus entailed a number of studies in the class voting literature. In a "new form of class voting" \nocite{oesch2018electoral} (Oesch and Rennwald 2018: 786), radical right parties receive disproportionate support from the manual working class, whereas "New Left" parties, such as Green parties, are more strongly supported by an emerging class of socio-cultural professionals \cite{kriesi2008west, bornschier2010new, bornschier2021us,rydgren2013class}. Whereas the Radical Right thus stands in close competition with the traditional center left, mainly social democratic parties, over the working class vote, its target groups also overlap with centre right voters, e.g. concerning small business owners in particular and the lower middle class in general (Oesch and Rennwald 2018). Next to class, educational attainment has the potential to divide the electorate into antagonistic groups. Stubager \citeyear{stubager2009education} could show that educational groups perceive group identity and inter-group conflict and differ crucially with regard to authoritarian and libertarian values. Strong rural place-based identities also seem to be more prominent among radical right voters \cite{zollinger2024cleavage}. Not least, the Radical Right attracts an older and rather male electorate \cite{spierings2015gendering, stockemer2022men}.

Whereas these structural configurations broadly summarize the voter side of the picture, part of a party's group image is also its own statement on the identities it represents. Parties compete not only via issues and positions, but also via emphasizing their ties (or antagonisms) to certain social groups. This serves parties' vote-seeking goals, because for identities to be relevant for (voting) behaviour, identity categories and resulting group-based conflicts need to be salient in voters' minds \cite{thau2023group, huddy2013group}. Identities thus need to be politicized \cite{bornschier2021us} in order to structure voter behaviour. This is why parties have crucial agency in shaping the developments outlined above, i.e., on which social identity voters base their decision to vote for the party. Naturally, parties' policy offer is one of the prime sources of political decision-making (see e.g. Evans and Tilley 2012)\nocite{evans2012depoliticization}. Group cues serve as an additional shortcut for making vote decisions \cite{dickson2006social}: they establish connections between parties and groups in voters' minds and if these groups are salient to them, help connect their own social identities to parties' group images.

Several recent studies have explored the group appeal strategies of single parties or party families: Thau \citeyear{thau2019political, thau2021social} could show how the Social Democrats in Denmark and the UK strategically broadened their group appeal to stray from their working-class image and tap into new voter bases. Huber (2022) as well as Stückelberger and Tresch (2022) show that parties appeal positively to those groups that enjoy high support in their electorate. The Radical Right's group appeal strategy has two main particularities: first, radical right parties put a lot of emphasis on out-group appeals \cite{heinisch2019populist, harteveld2022affective}. Mierke-Zatwarnicki \citeyear{mierke2023varieties} refers to this strategy as "oppositional identity politics" and traces it back to the unavailability of a sufficiently large and politicizable ingroup when radical right populists entered West European party systems. Second, Radical right parties often address low-status groups not (merely) via their sociostructural identities but via identities that offer identification with higher-status groups. Howe et al. \citeyear{howe2022nationalism} show how parties' appeals to "the nation" or co-nationals appealed to the declining agricultural sector of the working class in early 20th century Austria. Yet, it is unclear how particular the Radical Right's group image really is and how it influences other parties' group images. Do mainstream parties react to niche parties politicizing certain identities? In turn, do challenger parties broaden their repertoire of identity appeals over time, causing a convergence of mainstream and challenger parties' group images? We argue that this is the case and that the electoral success of RRPs as well as mainstream parties' vote loss foster this development. The next section will explore the theoretical underpinnings of this argument.


\section{A group perspective on mainstreaming niche and challenger parties}

Both theories of spatial and issue competition predict a "contagion" effect of (radical right) challenger parties: Especially centre left and centre right parties are expected to move their ideological position to the right in order to capture the challenger parties' voter potentials \cite{bale2003cinderella, van2010contagious, abou2016niche}. Several empirical findings yield support for this 'accomodative strategy' \cite{meguid2005competition, meguid2008party}, showing a positive effect of radical right vote share on mainstream parties' anti-immigration positions \cite{van2010contagious,han2015impact, abou2016niche,wagner2017radical}. With regard to issue competition, parties are expected to win from mobilizing an issue if it is salient in the campaign and the party possesses ownership in that issue \cite{budge1983explaining, budge2015issue}. Thus, especially the latter aspect incentivizes mainstream parties to increase their emphasis on the core issues of radical right challenger parties. Accordingly, empirical studies find an increase in salience of the immigration issue in response to the success of radical right parties \cite{green2019hot, gessler2022refugee}. Yet, two important considerations accompany this convergence of mainstream and radical right parties: first, while accommodation by mainstream parties is one side of the coin, moderation by radical right parties is the other \cite{wagner2017radical}. Not only can the Radical Right strategically moderate its positions \cite{wagner2017radical}, but also diversify its issue agenda \cite{bergman2020issue}. Second, competitiveness plays a crucial role for how strong the incentives for convergence are: with stronger radical right parties and mainstream parties facing electoral losses, convergence increases \cite{abou2016niche}.

We argue that we can expect similar effects for parties' group images. As outlined above, parties rhetorical group appeals serve to link the party with certain social group identities in voters' minds in the sense of a symbolic "standing for" kind of representation \cite{pitkin1967concept}. If successfully established, notions of "us" and "them" are related to voting behaviour \cite{bornschier2021us}, durably linking parties to social groups. Accordingly, mainstream parties have an incentive to challenge their niche competitors' group image, since they want to prevent the stabilization of group-based party-voter linkages in their favour. That is because identity-based links to groups of voters are generally favourable for parties, giving them a "valence advantage" \cite{dickson2006social} which results in less vulnerability in issue competition. Yet, similar to accommodation in issue salience, changing one's group image as a reaction to other parties' group image does not come without costs. There is continuous debate, for instance, about whether the traditional blue-collar working class is alienated by the Centre Left's inclusion of new sociodemographic groups into their appeal \cite{abou2021left}. Thus, we expect that mainstream parties' and radical right parties' group appeals converge when there is heightened competition, usually caused by the radical right party gaining electoral support. Then, the costs of not confronting the competitors' group image outweigh the risks of changing one's own group appeal. Thus we expect:
\vspace{2mm} 

\textbf{H1:} \textit{Increasing radical right party support increases the similarity between the Radical Rights' group image and the group image of its mainstream competitors.}

\vspace{5mm} 

Secondly, those parties that have lost votes in the previous election are more prone to consider changing their appeal strategy, an established finding in party competition literature \cite{adams2012causes}. Thus, if a Centre party has lost votes, it is more likely to try and speak to similar voter bases as its competitor(s). We there assume the following:
\vspace{2mm} 

\textbf{H2:} \textit{The more votes a mainstream party has lost at a previous election, the higher the similarity between the Radical Rights' group image and the group image of its mainstream competitors. }

\vspace{5mm} 

We expect that Radical Right support and Mainstream party vote loss also have a combined effect, because a party that has lost votes should have even more incentives to adapt to a competing party if the latter has gained votes - not only because the latter indicates a succesfull group appeal strategy, but also because it is then likely that the party's voters, at least in parts, have moved to the competitor. This is why we expect the interaction:
\vspace{2mm} 

\textbf{H3:} \textit{The stronger the Radical Right party, the more will vote loss of the Mainstream Party increase similarity between the Radical Rights' group image and the group image of its mainstream competitors.}

\vspace{5mm} 

 Centre Right parties are more likely to accommodate the Radical Right \cite{bale2003cinderella}, not only since they are potential coalition partners, but also, because of less divergent policies they are more likely to address similar social identities. However, we need to bear in mind that the Radical Right competes both with the Centre Left and with the Centre Right for specific social groups: with the Centre Left for the manual working class, and with the Centre Right for small business owners, for instance \cite{oesch2018electoral}. We therefore do not expect to see an effect of the right-left or GAL-TAN position of Centre parties on how similar their group image is to their radical right competitor.


\section{Research Design}

\subsection{Case selection}

To test our hypotheses, we use manifestos from 15 European countries in which Centre parties faced notable competition from Radical Right parties between 1980 and 2021. For a first test, we select the major Centre Left, the major Centre Right and the Major Radical right party by subsetting the Manifesto Dataset (Lehmann et al. 2023) to include only the Social Democratic (Centre Left), the Conservative and the Christian Democratic (Centre Right) and the Nationalist (Radical Right) party families. For each election, we include only the strongest (in terms of vote share) competitor of each party family. The sample is further restricted by excluding those radical right parties that never gained parliamentary seats over the period of observation and those parties that were in coalition with one another and published common manifestos (10 cases). In general, only those elections in which there was a radical right competitor are included, since the dependent variable is a dyadic measure of similarity between Centre and Radical Right. The final dataset thus includes two observations per election: the Centre Right and the Centre Left, including the calculation of their respective similarity to the Radical Right in that election. This leaves us with 35 Centre parties over 2 to 12 elections per country, amounting to 186 observations overall (see Table 1).

\begin{table}[htbp]
\footnotesize
\centering
\caption{Number of elections included per party family}
\begin{tabular}{lcc}
\toprule
Country & Centre Left & Centre Right \\
\midrule
Austria      & 12 & 12 \\
Belgium      & 9  & 10 \\
Denmark      & 12 & 12 \\
Finland      & 8  & 8  \\
France       & 4  & 4  \\
Germany      & 3  & 3  \\
Greece       & 4  & 5  \\
Hungary      & 3  & 3  \\
Italy        & 2  & 4  \\
Netherlands  & 7  & 7  \\
Norway       & 10 & 10 \\
Slovakia     & 4  & 4  \\
Spain        & 2  & 2  \\
Sweden       & 3  & 3  \\
Switzerland  & 8  & 8  \\
\bottomrule
\end{tabular}
\end{table}

Table 2 specifies the parties included in the dataset per country, i.e. the electorally strongest parties per party family in the respective year.\footnote{Note that strictly, the distinction between Centre parties are established, governing parties - in Switzerland for instance, the radical right SVP has mainstream status whereas liberal and conservative parties are its electorally weaker competitors. That is why we operate with "Centre" here and aim to perform different case selection rounds based on Challenger status \cite{vries2020political} and a nicheness measure instead of a binary classification \cite{bischof2017towards,meyer2013niche} for robustness.}

\begin{tiny}
	\begin{longtable}{|p{2cm}|p{3cm}|p{3cm}|p{3cm}|}
		\caption{Parties included per Country and Party Family}\\
		\hline
		\textbf{Country} & \textbf{Major mainstream left} & \textbf{Major mainstream right} & \textbf{Major radical right} \\
		\hline
		\hline
		Austria & Austrian Social Democratic Party (1983-2019)  & Austrian People’s Party (1983-2019)  & Austrian Freedom Party (1983-2019) \\
		Belgium & Francophone Socialist Party (1981-2014) & Christian Democratic and Flemish (1981-2019) & Flemish Bloc (1981, 1987-2003), Flemish Interest (2007-2019) \\
		Denmark & Social Democratic Party (1981-2019) & Conservative People’s Party (1981-2019) & Danish People’s Party (1998-2019), Progress Party (1981-1987, 1990-1994) \\
		Finland & Finnish Social Democrats (1983-2019) & National Coalition (1983-2019) & True Finns (1983-1991, 2003-2019) \\
		France & Socialist Party (1981-2002, 2012-2017) & Gaullists/Conservatives (1981-1988), Rally for the Republic (1993-1997), The Republicans (2002, 2012-2017) & National Front (1997-2017) \\
		Germany & Social Democratic Party of Germany (1980-2021) & Christian Democratic Union/Christian Social Union (1980-2021) & Alternative for Germany (2013-2021) \\
		Greece & Panhellenic Socialist Movement (2004-2015), The River (2015) & New Democracy (2004-2019) & Golden Dawn (2015-2019), Independent Greeks (2012), Popular Orthodox Rally (2007) \\
		Hungary & Hungarian Socialist Party (2002-2018) & Alliance of Federation of Young Democrats - Christian Democratic People's Party (2006-2018) & Movement for a Better Hungary (2010, 2014, 2018) \\
		Italy & Democratic Party (2008-2018), Italian Socialist Party (1983-1992) & Go Italy (1994-1996, 2018), People of Freedom (2008), Union of the Center (2013) & National Alliance (1983-1996), Northern League/League (2008, 2018) \\
		Netherlands & Labour Party (1982-2021) & Christian Democratic Appeal (1982-2002, 2006-2021) & Centre Democrats (1989-1994), List Pim Fortuyn (2003), Party of Freedom (2006-2021) \\
		Norway & Labour Party (1981-2017) & Conservative Party (1981-2017) & Progress Party (1981-2017) \\
		Slovakia & Direction-Social Democracy (2002-2016), Party of the Democratic Left (1992) & Christian Democratic Movement (1990, 1994, 2012), Ordinary People and Independent Personalities (2016), Slovak Democratic and Christian Union (2002-2010) & Slovak National Party (2006-2016) \\
		Spain & Spanish Socialist Workers’ Party (1982-2019) & People's Party (1982-2019) & Voice (2019) \\
		Sweden & Social Democratic Labour Party (1982-2018) & Moderate Coalition Party (1982-2018) & Sweden Democrats (2010-2018) \\
		Switzerland & Social Democratic Party of Switzerland (1983-2019) & Christian Democratic People’s Party of Switzerland (1983-2019) & Swiss People’s Party (1983-2019) \\
	\end{longtable}
\end{tiny}

\subsection{Computation of Social Group Labels}

\subsubsection{Dictionary Approach}
For the first identification of group appeals in manifesto sentences, we produced a dictionary containing references to 22 social groups with several synonyms for each group. The dictionary is taken from Riethmüller \citeyear{riethmueller2024thecontestaion} and restricts the sample of groups to those groups with a clear sociodemographic basis, which we determined by using those groups that are contained in the demographic variables of the \textit{Comparative Study of Electoral Systems (CSES)}, Waves 1-5: age (young people, older people), gender (women, men), education level (high, low), religion (Christians, Muslims)\footnote{As the original dictionary was reduced to groups for which the n of respondents in Germany, Austria, and Switzerland per CSES wave was high enough, other religions are excluded here.}, income groups (high, low), unemployed people, place of living (rural, urban), migration history, parents and families, occupation-based groups (manual workers, care workers, farmers, academic professionals, soldiers), students, entrepreneurs. This does not only allow future compatibility with demand-side data but also creates a whitelist of seed words/seed groups that has clear boundaries. This represents a rather narrow conceptualization of what a social group is. This whitelist can then serve as a basis for expanding the identification to groups not previously defined by the researcher.

\subsubsection{BERT-based Classification of Social Group Labels}

The group labels created with the dictionary approach with the German manifestos from the Manifesto Project (Lehmann et al. 2023) were used as a training set for training a BERT model \cite{devleretal2018BERT}. The expectation was that BERT would find similar sentences to the labeled ones and thus extend the sensibility of the dictionary approach to different expressions and languages. 

Although a multilingual base model was used the labels did not translate well to other languages. Languages (or countries) that either held an representation bias in the BERT model or were very different from German (Greek, Hungarian, Finnish) displayed the worst recall. Due to this, the entries were translated to German using DeepL \cite{gtranslate2020}. Subsequently, the German dictionary was used to label a training set for these languages and retrain the model. 

The model was trained using 10 epochs and an Adams Optimizer \cite{Kingma2014ADAM}.  The eval f1 score was 0.98.
On the test data, the following results were obtained: 

\begin{table}[h!]
\centering
\caption{Performance Metrics}
\begin{tabular}{@{}lccc@{}}
\toprule
Group            & Precision & Recall & F-score \\ \midrule
No Group         & 0.82      & 0.91   & 0.86    \\
Multiple Groups  & 0.76      & 0.63   & 0.69    \\
Overall          & 0.98      & 0.99   & 0.99    \\ \bottomrule
\end{tabular}
\end{table}

Although the BERT model successfully reproduced the social group labels the yield of attaching labels to cases the dictionary missed was not as high as expected. This was due to sampling problems: it was not clear which cases should be sampled as not containing a group without manual investigation. 

\subsubsection{Combined approach Mistral-LLM(LLM) and Embedding Space-based Filtering (ESF)}

Due to limitations of the BERT-based approach state-of-the-art transformer models were explored as an alternative approach. Instead of fine-tuning these models, they were used for a one-shot inference to label the social groups. Manual investigation showed that these LLMs are sufficiently trained to detect social groups given the correct prompt. Google Gemini \cite{geminiteam2024gemini}, Metas LLama \cite{touvron2023llama}, ChatGPT and HuggingFace's Mistral \cite{mistralai2024} were used. 

A quantized version of Mistral-7B-Instruct-v0.1 was used on Google Colab to produce the social group labels asking for both explicit and implicit social groups contained in the text. The request for implicit groups was used to help the model understand the concept of explicit, which is what we are looking for, better and sort the groups where it is unsure into "implicit" (see e.g. Zhuang et al. 2023)\nocite{zhuang2023beyond}. In the prompt, we included a minimal definition of what a social group is, based on Licht and Sczepanski (2023, p.12) The following prompt was used to generate the social group labels: 

\bigskip
\textsl{[INST] Eine soziale Gruppe ist eine Gruppe von Personen mit einem gemeinsamen Merkmal.
Nenne die sozialen Gruppen in dieser Aussage, die explizit und direkt sprachlich genannt werden.
Nenne auch die Gruppen, um die es implizit geht. Wenn es andere Nomen oder Begriffe gibt,
nenne sie "Sonstige". Gibt mir den Output bitte als json Format und keinen weiteren Text.
Die Aussage ist: \{\}[/INST]}

\bigskip
This translates to:

\bigskip
\textsl{[INST] A social group is a group of people with a common characteristic. Name the social groups explicitly and directly mentioned in this statement. Also, name the groups that are implied. If there are other nouns or terms, call them "Others". Please provide the output in JSON format and no additional text. The statement is: \{\} [/INST]
}

\bigskip
Although the results found a promising number of social groups, greatly enhancing the dictionary, the output was noisy and needed filtering. Using the BERT embeddings of the found groups and those of the dictionary, a classifier was developed capable of filtering those groups semantically close to the seed list.  The assumption here is that the embedding of the label \textbf{Social Group} itself is close to the geometric center of all the embeddings representing the dictionary.

\bigskip
\begin{figure}[ht!]
    \centering
    \includegraphics[width=\textwidth]{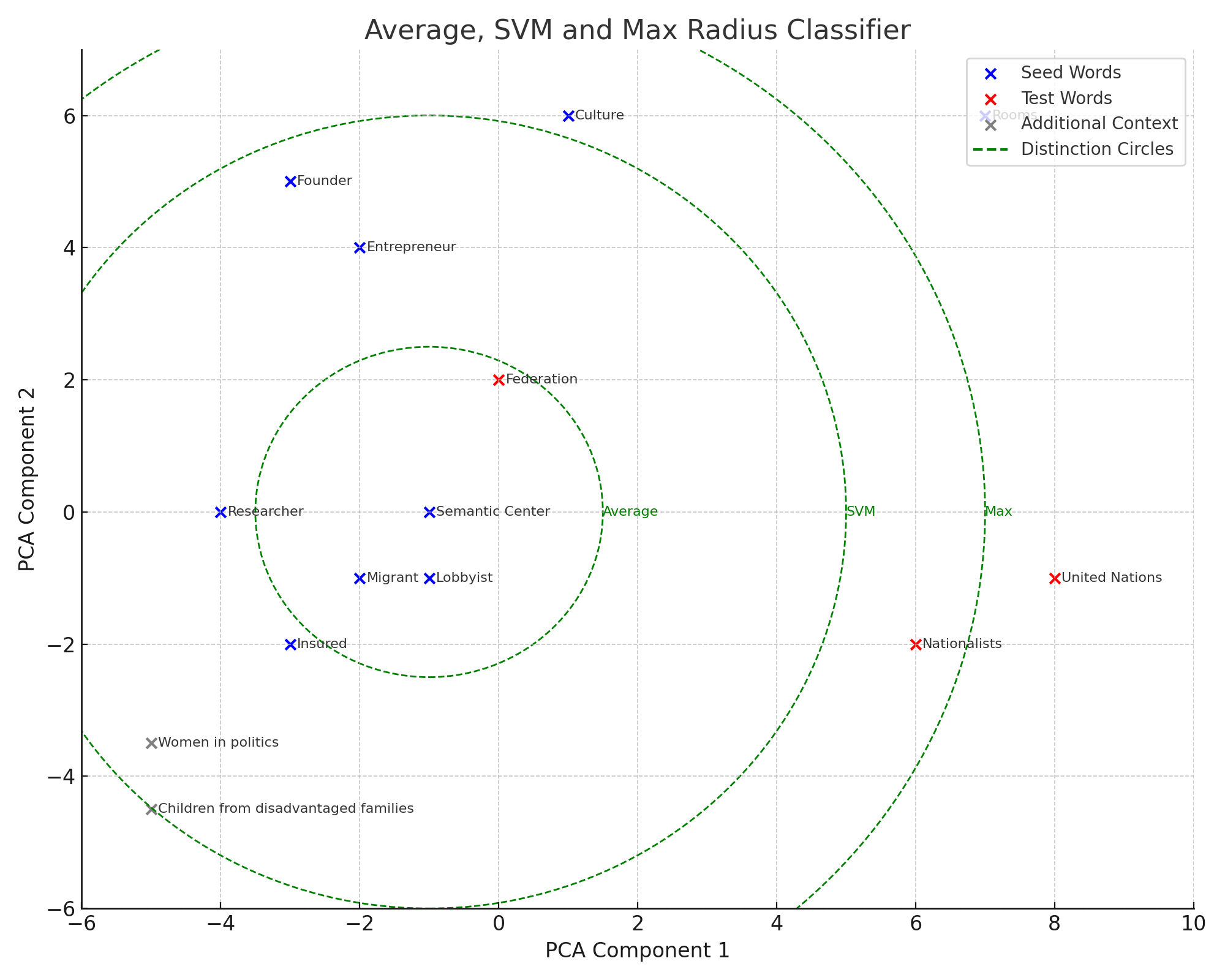}
    \caption{Visualization of the Embedding Space-based Filtering (ESF)}
    \label{fig:enter-label}
\end{figure}

In figure \ref{fig:enter-label} a simplified version of the algorithm with exemplary group labels is displayed. For the figure, the dimensions were reduced to two using PCA. Even in this simplified representation of the words, concepts that depict social groups like Migrant, Insured, or Women in Politics are closer to the Semantic Center than concepts like the United Nations. The Semantic Center is the geometric point of the average of all the embeddings of the dictionary (a white list for what constitutes a social group). 
The distinction circles represent the three classifiers that were used. The most conservative one is the average of all the embeddings of the whitelist. The outer line is the maximum distance of an embedding in the whitelist to the Semantic Center. Using the same data a OneClass SVM model was trained. It optimizes the hyper-sphere to shape itself to the data. The distinction lines are less a circle in higher dimensions but an n-dimensional ellipsoid. Training the perfect classifier this way takes several iterations. The conservative average line classification was used because there were still too many false classifications. 

The three classification approaches based on the proximity to a "Semantic Center" are detailed below:

\subsection*{1. Average Distance Classifier}
The average distance classifier computes the average of all embeddings in the whitelist (dictionary) and uses this average as a reference point for classification. The assumption is that the mean embedding represents the semantic center of social groups. In the description of the classifier, only the distinction line is detailed. The algorithm separating the words inside and outside the distinction line is trivial.

\textbf{Formula:}
\begin{equation}
\mathbf{c}_{\text{avg}} = \frac{1}{N} \sum_{i=1}^N \mathbf{x}_i
\end{equation}
Where:
\begin{itemize}
    \item $\mathbf{c}_{\text{avg}}$ is the average (or mean) embedding vector.
    \item $N$ is the number of embeddings in the whitelist.
    \item $\mathbf{x}_i$ is the embedding of the $i$-th entry in the whitelist.
\end{itemize}

\subsection*{2. Maximum Distance Classifier}
The maximum distance classifier involves computing the distance from each whitelist embedding to the semantic center and then determining the maximum of these distances. This approach identifies the furthest point in the whitelist embeddings from the center and uses this distance to form a boundary.

\textbf{Formula:}
\begin{equation}
d_{\text{max}} = \max_{i=1}^N \|\mathbf{x}_i - \mathbf{c}_{\text{avg}}\|
\end{equation}
Where:
\begin{itemize}
    \item $d_{\text{max}}$ is the maximum distance from the semantic center.
    \item $\|\cdot\|$ denotes the Euclidean norm.
\end{itemize}

\subsection*{3. One-Class SVM Classifier}
The One-Class SVM~\cite{Bounsiar6847442} approach attempts to fit an n-dimensional ellipsoid around the data points, optimizing the boundaries to include the group of embeddings that represent the concept of social groups most closely.

\textbf{Formula:}
The general form of a One-Class SVM problem can be expressed as:
\begin{equation}
\min_{\mathbf{w}, \xi_i, \rho} \frac{1}{2} \|\mathbf{w}\|^2 + \frac{1}{\nu N} \sum_{i=1}^N \xi_i - \rho
\end{equation}
Subject to:
\begin{equation}
\mathbf{w} \cdot \phi(\mathbf{x}_i) \geq \rho - \xi_i, \quad \xi_i \geq 0, \quad \text{for all } i
\end{equation}
Where:
\begin{itemize}
    \item $\mathbf{w}$ is the weight vector of the hyperplane.
    \item $\xi_i$ are slack variables allowing for some points to be outside the decision boundary.
    \item $\rho$ is the offset of the decision boundary from the origin in the feature space.
    \item $\nu$ is an SVM parameter that controls the trade-off between maximizing the decision function's margin and minimizing the fraction of outliers.
    \item $\phi(\mathbf{x}_i)$ is the feature map transforming data into a higher dimensional space.
\end{itemize}

\subsection{Social Group Detection Approaches}

With the LLM-ESF procedure revealing a number of additional groups to those in the seed dictionary, we decided to enhance the seed dictionary with the LLM-ESF-identified social groups, to make sure they are being found in every instance.
To compare the three approaches, we hand-coded a sample of 840 sentences, stratified by language, according to whether it contains a social group (one of the seed groups or any group beyond the seed groups). We relied on the automated translation for this step, which needs more careful validation in future iterations of this paper. We found that whereas both the dictionary and the LLM-ESF approach perform quite well (in terms of the Micro F1 Score), there is a considerable set of social groups in manifesto sentences that the seed dictionary misses and that the LLM-ESF is better at capturing.

\begin{table}[ht]
\centering
\caption{Summary of F1 Scores by Model}
\label{tab:model_performance}
\begin{tabular}{@{}lcc@{}}
\toprule
\textbf{Model} & \textbf{Micro F1 Score} & \textbf{Macro F1 Score} \\
\midrule
Dictionary & 0.83 & 0.57 \\
BERT & 0.53 & 0.21 \\
LLM-ESF & 0.89 & 0.62 \\
\bottomrule
\end{tabular}
\end{table}

The BERT model likely underperformed  for various reasons

\begin{itemize}
    \item oversampling of dictionary groups in training sample
    \item missing goldstandard data in case no group was found
    \item language or context bias of groups
\end{itemize}

Using the combination of LLM and ESF (what we call LLM-ESF for short) yielded the best result and produced an equally stable automated pipeline as the BERT-model but with better results and a number of newly found groups. In contrast to the other approaches it finds all existing groups and needs to be fine-tuned to reduce noise. Even though it still requires human coding iterations, it scales quickly with each group added to the dictionary. The approach also generalizes to other CSS applications.

\subsection{Operationalization}

\paragraph{\textit{Dependent variable:}} We construct a measure of similarity between Centre Right/Left 
and Radical Right parties' group appeals that is based on a salience concept of social group appeals: 
we identify relevant appeals to social groups using the LLM-ESF classification-based on the sentence-level and then
calculate each groups' salience per party manifesto by dividing the number of sentences containing a reference to this
group by the number of all sentences in the manifesto that contain a reference to a social group. We therefore use the Manifesto Corpus (Lehmann et al. 2023) \nocite{lehmann_et_al_2023}  and split the fulltexts into natural sentences. For these salience values between 0 and 1
we can then calculate a similarity index based on the index of programmatic dissimilarity \cite{hennl_franzmann_2017}. For two parties, it is calculated by subtracting the salience that
party A ascribes to this group from the salience that party B ascribes to this group, sums up these differences and divides the result by two. This results in a measure ranging from 0 to 100, where 100 means perfect similarity in two parties' group images. The salience approach to social group appeals does not only refer to the relative calculation, but we also do not distinguish between positive and negative/in-group vs. out-group appeals here.

\paragraph{\textit{Independent variables}}

The data for the \textit{Radical Right's vote share} comes from the Manifesto Project' database (Lehmann et al. 2023) \nocite{Lehmann2023}. We include a lagged version of this variable to test the influence of the Radical Right's vote share at election t-1. We calculate the \textit{Centre parties' vote difference} at t-1 by subtracting its vote share at election t-1 from its vote share at election t-2. A negative value thus means the party has lost votes in the previous election. Further, similar to \cite{abou2016niche}, we control for whether the Centre party was in government in the period before the manifesto and the party's vote share at t. The data on government periods are taken from the PPEG database \cite{krause_stelzle_2024}.

\begin{table}[htbp]
  \centering
  \caption{Descriptive statistics}
  \begin{tabular}{@{}lccccc@{}}
    \toprule
    Statistic   & N   & Mean  & St. Dev. & Min  & Max  \\ 
    \midrule
    Similarity \textbf{(DV)}  & 186 & 39.6  & 17.5     & 0.0  & 74.4 \\
    Radical Right voteshare       & 186 & 12.3  & 7.0      & 0.9  & 29.4 \\
   Mainstream party vote difference & 175 & -0.9  & 6.9      & -32.9& 25.4 \\
    Gov. party  & 186 & 0.7   & 0.5      & 0    & 1    \\
    Mainstream vote share     & 186 & 23.5  & 10.4     & 3.4  & 67.9 \\
    \bottomrule
  \end{tabular}
\end{table}

\subsection{Modelling strategy}
To test our hypotheses, we estimate time-series cross-section regression models based on OLS regressions, since our dependent variable (similarity) is continuous. We calculate election-clustered standard errors to mitigate panel-specific heteroskedasticity. Similar to Abou-Chadi (2016), we include party-dummy fixed-effects. This allows us to exclude time invariant country- or party level factors from our analysis and focus solely on within-party variation over time.

\section{Descriptive results}

\begin{figure}[h]
    \centering
    \includegraphics[width=1.1\linewidth]{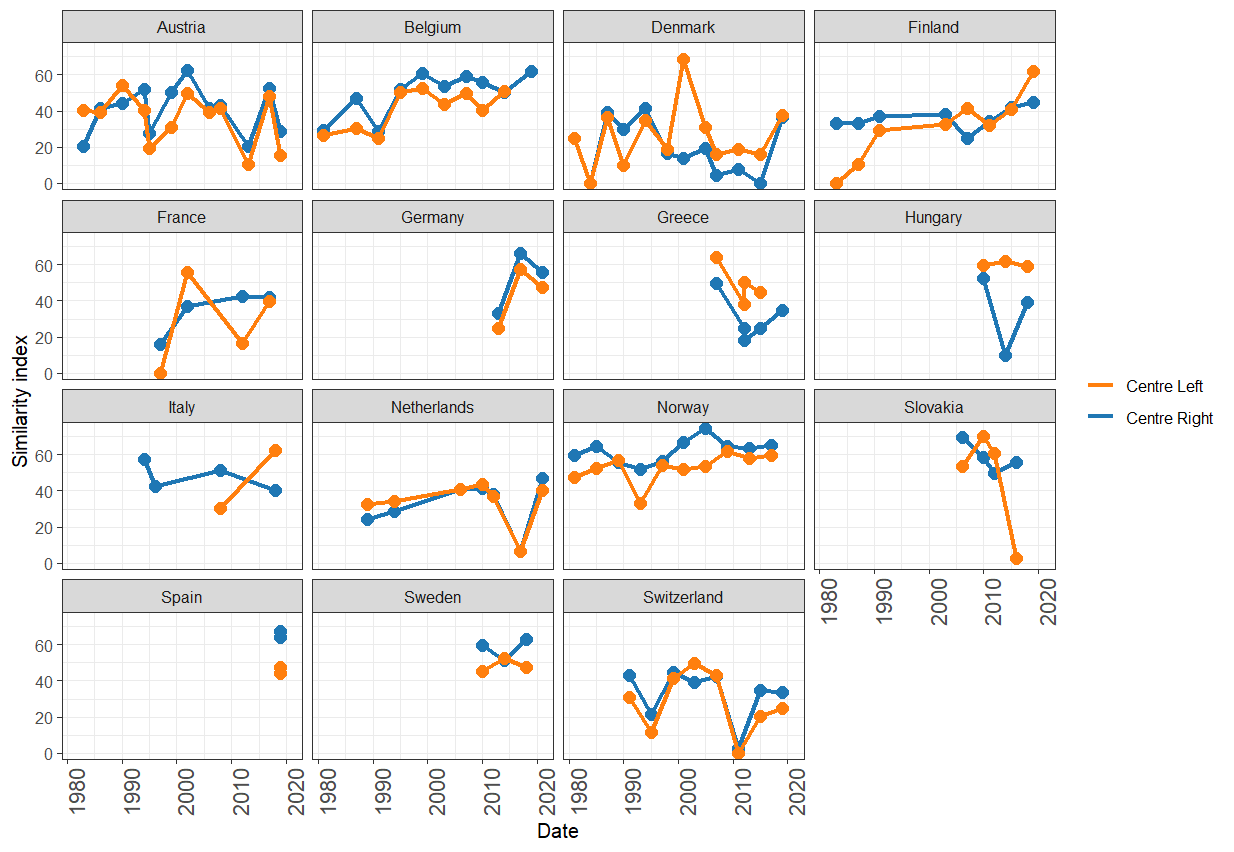}
    \caption{Similarity (0-100) to the Radical Right's group appeal per election. The calculation is based on electoral manifestos (Lehmann et al. 2023a).}
    \label{fig:2}
\end{figure}

Figure \ref{fig:2} shows the overlap between the Centre Left's and the Centre Right's group appeals or group image with those of the Radical Right respectively per country and election. A value of 100 means that the party appeals to exactly the same groups as the Radical Right to the exact same extent. A value of zero means there is no overlap between the groups the Radical Right appeals to in their manifesto and those the respective mainstream party appeals to. Several aspects are notable: Firstly, we do not see large differences between the Centre Left and the Centre Right in terms of their similarity in group appeals to the Radical Right. Whereas in some countries, such as Germany, Norway, and Belgium, the group image of the Centre Right overlaps more closely with the Radical Right over the whole observation period, in Greece for instance the opposite is the case and in Italy, Denmark, and Finland the Radical Right - Centre Left similarity surpasses Radical-Right - Centre Right similarity eventually. Second, some outliers are striking, in which very high or very low similarity scores are reached. Understanding how those come about is useful for understanding group image similarity more generally.In the Netherlands 2017 for instance, the radical right PVV competed only with one-page program in which the only groups addressed were immigrants, asylum seekers, muslims, and (Dutch) citizens. The Centre Left (PvdA) and the Centre Right (CDA) addressed these groups only marginally, none of them addressed muslims for instance. They instead focused on a broader range of other social groups, leading to an overlap close to zero.

\begin{figure}
    \centering
    \begin{subfigure}{0.5\linewidth} 
        \centering
        \includegraphics[width=\linewidth]{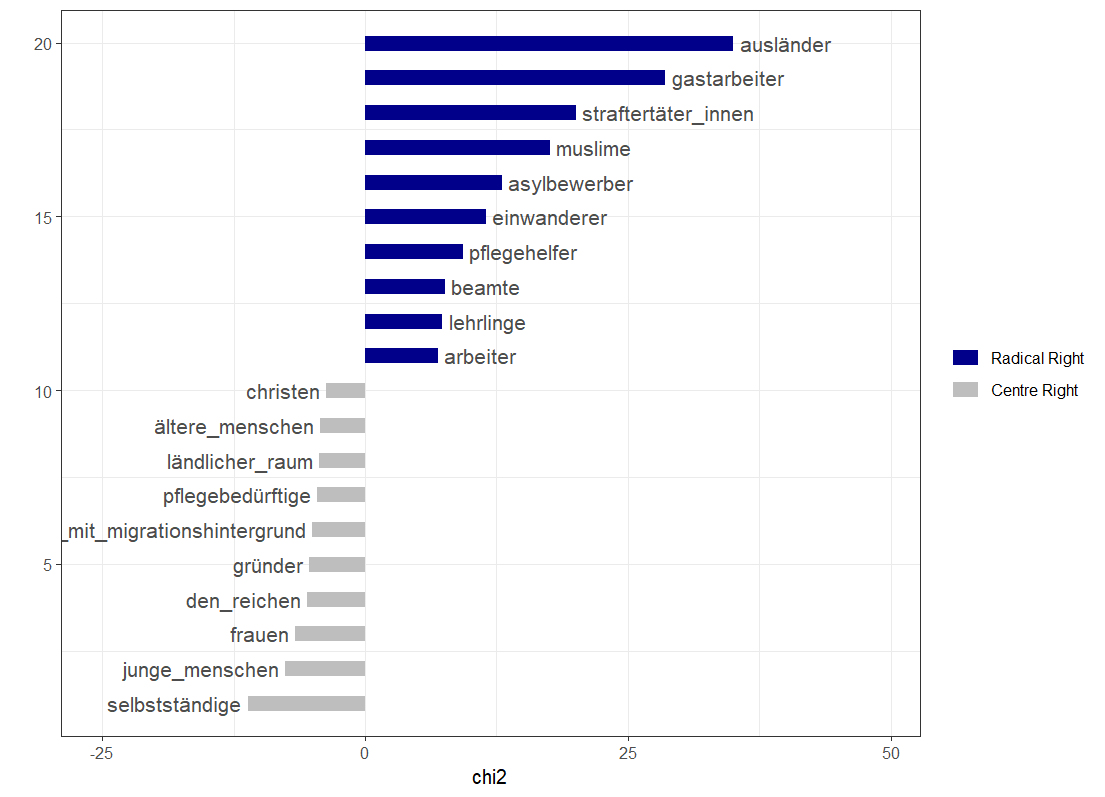}
        \caption{Keyness Radical Right vs. Centre Right}
        \label{fig:subfig1}
    \end{subfigure}%
    \begin{subfigure}{0.5\linewidth} 
        \centering
        \includegraphics[width=\linewidth]{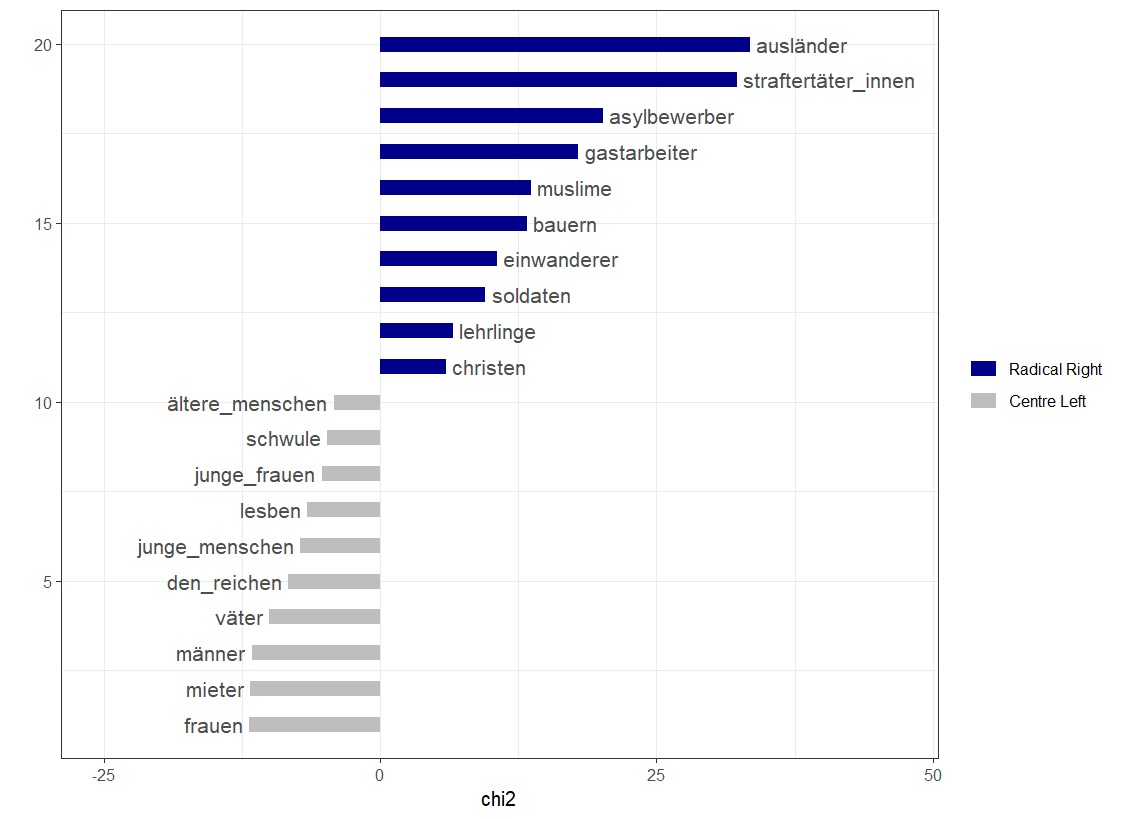}
        \caption{Keyness Radical Right vs. Centre Left}
        \label{fig:subfig2}
    \end{subfigure}
    \caption{Groups mentioned with highest relative frequency by Radical Right compared to Centre Left Parties. Translation for a, from above: \textit{foreigners, guest workers, criminals, Muslims, asylum seekers, immigrants, nursing assistants, civil servants, apprentices, (manual) workers, Christians, older people, rural areas, those in need of care, people with migration backgrounds, start-up founders, the rich, women, young people, self-employed}. Translation for b, from above: \textit{foreigners, criminals, asylum seekers, guest workers, Muslims, farmers, immigrants, soldiers, apprentices, Christians, older people, gay people, young women, lesbian people, young people, the rich, fathers, men, tenants, women}}
    \label{fig:overall}
\end{figure}

An advantage of our approach in identifying all social group appeals in their various forms lies in the possibility to compare \textit{how} different parties and party families talk about social groups and how they construct their group image. To illustrate how they differ, we can compare each party family's group image using keyness statistics, which identifies which words are most "typical" or distinctive for a text, based on significant differences in relative frequencies. Since we are interested in overlap between the Radical Right with the two other main party poles \cite{zollinger2024cleavage}, this is very telling, since it taps precisely into the concept of "group image": the social groups a party positively or negatively links itself to \textit{in contrast to other parties}.

Not surprisingly, the Radical Right is most distinct in their frequent addresses of foreigners, immigrants and asylum seekers. Here, we see a difference between Centre Right and Radical Right parties worth noting: the Centre Right does not avoid migration-related groups, but appeals to "people with migration backgrounds" more frequently than the Radical Right, indicating a slight difference in how this group is framed. Whereas these groups could all have been summarized under "migrants" in other approaches, an advantage of our approach is that exactly these differences in appeals become visible. Regarding the  Radical Right's class appeal, it is further interesting to see that the Radical Right's group appeal is distinct from the Centre Right's group appeal in their emphasis on workers, whereas they do not appeal to workers significantly more or less than the Centre Left. Integrating subfigures a and b, the Radical Right's distinctive group image emerges as out-group centered and centered around low-status or status-anxious demographics like nurses, apprentices, or farmers.

\section{Effects}
Table 6 presents the results of the three regression models. It is visible that the Radical Right's electoral support in the previous election does not have a statistically significant impact on the similarity of Mainstream and Radical Right's group appeals. Vote loss by the Center parties does not seem to induce an increasing overlap with the Radical Right's group appeal either, which is why we do not find support for hypotheses H1 to H3 here. Model 2 and 3, however, find that government parties are indeed less close to Radical Right parties in which groups they address. This might be due to these parties having less room to maneuver since there are institutionalized expectations about which social groups a governing party ought to address. This could be an interesting direction for further explorations.

The results indicate that the variance in similarity between Mainstream and Radical Right parties cannot be explained by a contagion effect of Radical Right parties. Mainstream parties do not seem to adapt their group images to those of a competing Radical Right party as a response to electoral pressure. We do see however increasing similarity in many countries under study, especially those in which there is a long-established Radical Right party, such as Belgium, Finland, Norway, or the Netherlands (see Figure 2). Thus, there might be other factors explaining whether parties do or do not appeal to a similar set of social groups which we have not explored here yet.

\begin{table}[!htbp] \centering 
  \caption{Results of pooled OLS regression models (cluster-robust standard errors in parentheses)} 
  \label{} 
\begin{tabular}{@{\extracolsep{5pt}}lccc} 
\\[-1.8ex]\hline 
\hline \\[-1.8ex] 
\\[-1.8ex] & (1) & (2) & (3)\\ 
\hline \\[-1.8ex] 
 Radical Right support (t-1) & 0.198 &  & 0.291 \\ 
  & (0.236) &  & (0.236) \\ 
  & & & \\ 
 Vote difference Centre &  & 0.131 & 0.194 \\ 
  &  & (0.159) & (0.418) \\ 
  & & & \\ 
  RR support(t-1):Vote diff. Centre &  &  & $-$0.013 \\ 
  &  &  & (0.033) \\ 
  & & & \\ 
 Government party & $-$4.837 & $-$5.478$^{*}$ & $-$6.719$^{*}$ \\ 
  & (2.567) & (2.409) & (2.589) \\ 
  & & & \\ 
 Voteshare Centre & 0.455 & 0.304 & 0.662$^{*}$ \\ 
  & (0.256) & (0.207) & (0.269) \\ 
  & & & \\ 
 Constant & 39.328$^{**}$ & 43.197$^{***}$ & 34.486$^{**}$ \\ 
  & (12.608) & (9.955) & (12.776) \\ 
   Party fixed effects & \checkmark & \checkmark & \checkmark \\ 
  & & & \\ 
\hline \\[-1.8ex] 
Observations & 157 & 175 & 151 \\ 
R$^{2}$ & 0.515 & 0.485 & 0.537 \\ 
Adjusted R$^{2}$ & 0.370 & 0.365 & 0.397 \\ 
F Statistic & 3.540$^{***}$ & 4.025$^{***}$ & 3.816$^{***}$ \\ 
\hline 
\hline \\[-1.8ex] 
\textit{Note:}  & \multicolumn{3}{r}{$^{*}$p$<$0.05; $^{**}$p$<$0.01; $^{***}$p$<$0.001} \\ 
\end{tabular} 
\end{table}


\section{Conclusion and discussion}

In this study, we present a novel computational text classification strategy to identify references to social groups within political texts, specifically party manifestos: Our combination of LLM-based one-shot labeling in combination with Embedding Space-based Filtering \textit{(LLM-ESF)} makes it possible to start from a list of seed groups and automatically expand this list to ultimately all social groups appearing in a set of unseen texts. Thus, we address the challenges researchers face when a full picture of parties' group-based rhetoric is required, particularly in large-scale cross-national analyses. By integrating supervised identification using a predefined dictionary of social groups, training a BERT multilingual language model to discern semantic meanings and alternative expressions, and leveraging the Mistral LLM to extract explicitly mentioned social groups without prior definition, we arrive at a promising method of deriving social group appeals from political texts. 

The comprehensive set of social group appeals that the LLM-ESF-enhances identification yields allowed us to compare Radical Right parties' group images to those of their mainstream competitors. We could thus map how similar Radical Right parties and mainstream parties are in terms of which groups they address and test whether Radical Right electoral support and mainstream vote loss cause parties to move closer together in their group appeal. However, these factors cannot explain the variance in similarity of group appeals which we can clearly see. In future iterations of this paper, we will test different ways of measuring the dependent variable. A fruitful approach could for instance be to use embedding-based similarity measures, which are more fine-grained than our current category-based similarity index and could identify for instance that "young women" and "young families" are more similar than "young women" and "pensioners".

Our LLM-ESF approach to identifying social groups in large text corpora is of relevance to the literature studying parties' social group appeals \cite{huber2022beyond, dolinsky2023parties, stuckelberger2022group, thau2019political, thau2021social, thau2023group}, especially to extend analyses across time and space. It provides a valuable tool for scholars seeking the role of group identity in party competition and the supply side to identity-formation dynamics that have been analysed on the demand-side \cite{bornschier2021us, zollinger2024cleavage}. 

For future versions of this paper, we seek to validate our approach against hand-coded (e.g. Huber 2022; Thau 2019; Dolinsky 2023) data and other automated approaches (Licht and Szepanski 2023). Further, we will work on making the method suitable for multilingual applications without having to rely on automated translation. Moreover, continuous updates of the seed list should improve the accuracy in comparison to the gold standard making the SVM approach superior to the conservative average embedding classifier. Finally, the pipeline LLM-ESF will be made available open-source as a library. 
\newpage 


\bibliography{literature.bib} 
\bibliographystyle{apsr} 




\end{document}